# Identifying Plastic Inorganic Semiconductors Requires More Rigorous Criteria

Qiao Wang

School of Chemistry and Chemical Engineering, Harbin Institute of Technology, Harbin 150001, China

**Abstract**

The identification of plastic inorganic semiconductors becomes challenging when their mechanical responses depend on crystallographic orientation, sample size, and loading conditions. Using layered GeSe as a model system, we examine its deformation behavior through macroscopic compression, bending, conventional micropillar compression, and eccentric micropillar compression. Under macroscopic compression perpendicular to the layers, GeSe sustains approximately 23% strain without fracture, whereas bending along the c-axis armchair direction produces brittle cleavage fracture. Conventional micropillar compression results in brittle fragmentation, while eccentric loading introduces a shear component that activates pronounced interlayer sliding and accommodates deformation. These contrasting responses reflect the coupled effects of bonding topology, interlayer van der Waals interactions, defect density, stress constraints, and strain path on the competition between sliding and fracture. The results highlight the limitations of identifying plasticity from a single direction, scale, or test and support a systematic evaluation framework combining multiple directions, length scales, loading modes, and characterization techniques. More rigorous and unified criteria are needed to guide reliable materials selection for flexible electronics and devices integrated on curved surfaces.

## Main text

Since their emergence, plastic inorganic semiconductors have combined the advantages of organic and inorganic materials, retaining the superior electrical performance of inorganic semiconductors while offering mechanical flexibility similar to that of organic materials.[1] They thus provide a new materials approach for wearable devices and applications involving specially curved surfaces. As new plastic material systems and mechanisms continue to emerge,[2–4] particularly as research extends to materials exhibiting ductile–brittle anisotropy, whether these materials truly exhibit plasticity often becomes contentious and can be misjudged. Such materials may display markedly different mechanical behaviors depending on crystallographic orientation, loading direction, length scale, or stress state. Consequently, conventional conclusions of "plasticity" based on a single direction, test, or criterion may be unreliable. It is therefore essential to establish more comprehensive and systematic characterization methods and adopt more rigorous, or even stricter, evaluation criteria. These efforts will facilitate the accurate identification and elucidation of deformation mechanisms in plastic inorganic semiconductors and provide a solid basis for their reliable application in flexible electronics, wearable devices, and integration on specially curved surfaces.

Against this background, we use GeSe as a representative case to systematically investigate its plasticity and ductile–brittle behavior across different length scales, loading modes, and characterization techniques. GeSe and related Ge-based materials have broad application prospects in photovoltaics, photodetection, thermoelectric conversion, phase-change memory, lithium-ion battery anodes, and flexible electronics.[5–14] GeSe crystallizes in the orthorhombic system (space group *Pnma*), and its crystal structure is shown in Fig. 1a. Within each layer,

Ge and Se atoms are connected by covalent bonds, whereas adjacent layers are stacked through van der Waals interactions, giving rise to a characteristic layered structure and pronounced anisotropy. We synthesized GeSe crystals by chemical vapor transport (CVT) (Fig. 1b). The resulting crystals were grayish black, with a metallic luster and a distinct layered morphology. In the X-ray diffraction (XRD) pattern (Fig. 1c), the diffraction peaks at $2\theta$ = 16.56°, 33.27°, 50.71°, and 69.54° correspond to the (200), (400), (600), and (800) planes, respectively, of an (*h*00)-textured crystal, indicating the successful synthesis of layered GeSe crystals with a high degree of preferred orientation along the *a* axis. These high-quality crystals provide an ideal model system for subsequent characterization of plasticity and analysis of its mechanisms across multiple length scales, loading modes, and characterization techniques in the context of ductile–brittle anisotropy.

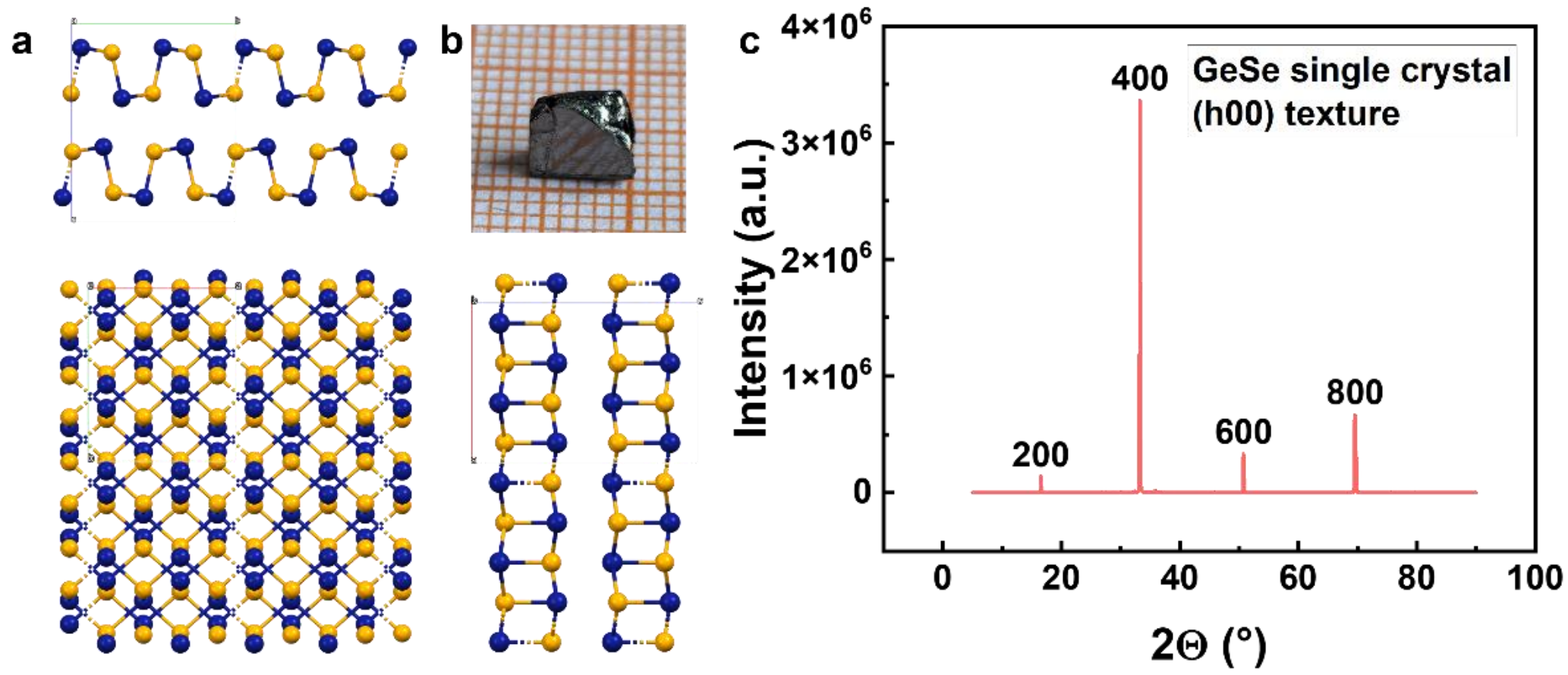

**Figure 1.** Structure and synthesis of GeSe crystals. (a) Schematic illustration of the GeSe crystal structure. (b) Photograph of the synthesized GeSe crystals. (c) XRD pattern of GeSe.

Figure 2 shows a typical stress–strain curve of a GeSe crystal under compressive loading. The arrow in the inset indicates that the loading direction is perpendicular to the plane of the layered structure. Initially, the stress rises rapidly with strain, and yielding begins at a strain of approximately 1%. Thereafter, the stress increases gradually with increasing strain, exhibiting pronounced continuous strain hardening. At a strain of approximately 23%, the stress exceeds 10 MPa, and the curve still shows no indication of fracture or a sudden load drop, demonstrating the excellent plastic deformation capacity of GeSe under macroscopic compression. Notably, the stress axis is logarithmic and spans three orders of magnitude, from an initial value of 0.01 MPa to over 10 MPa, reflecting a continuous increase in the load-bearing capacity during deformation. This large plastic deformation may be associated with the weak interlayer van der Waals interactions in the layered GeSe structure, which facilitate interlayer sliding; subsequent micropillar compression experiments confirm this. Furthermore, because GeSe is orthorhombic, its mechanical response is markedly anisotropic. According to the literature, GeSe is isostructural with SnSe and exhibits plasticity along the in-plane *b*-axis zigzag direction and brittleness along the in-plane *c*-axis armchair direction. The SEM image in Fig. 3 further shows that the fracture surface after bending along the *c* axis is flat and displays typical features of brittle fracture, with no obvious crack deflection or plastic deformation zone. These results indicate that the plasticity and ductile–brittle behavior of GeSe are strongly direction-dependent and differ substantially with the testing method.

Therefore, when evaluating plasticity in GeSe or other plastic inorganic semiconductors, the loading direction, sample size, and test conditions must be clearly specified. Conclusions cannot be drawn from a single direction or criterion alone, further emphasizing the importance of comprehensive, systematic characterization methods and stricter evaluation criteria.

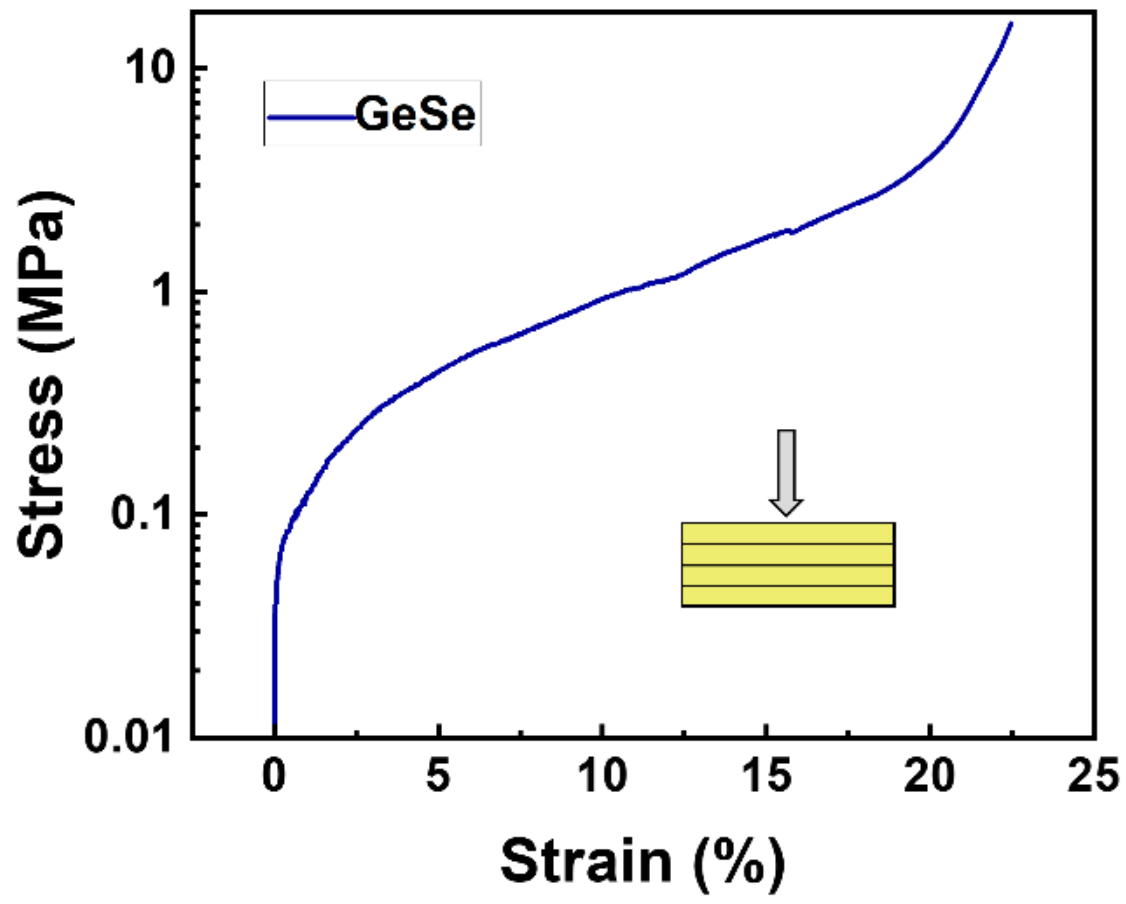

**Figure 2.** Compressive stress–strain curve of a GeSe crystal. The sample dimensions are 4.47 mm × 7.02 mm × 2.32 mm, and the compression rate is 0.5 mm/min. The inset shows the loading direction perpendicular to the plane of the layered structure.

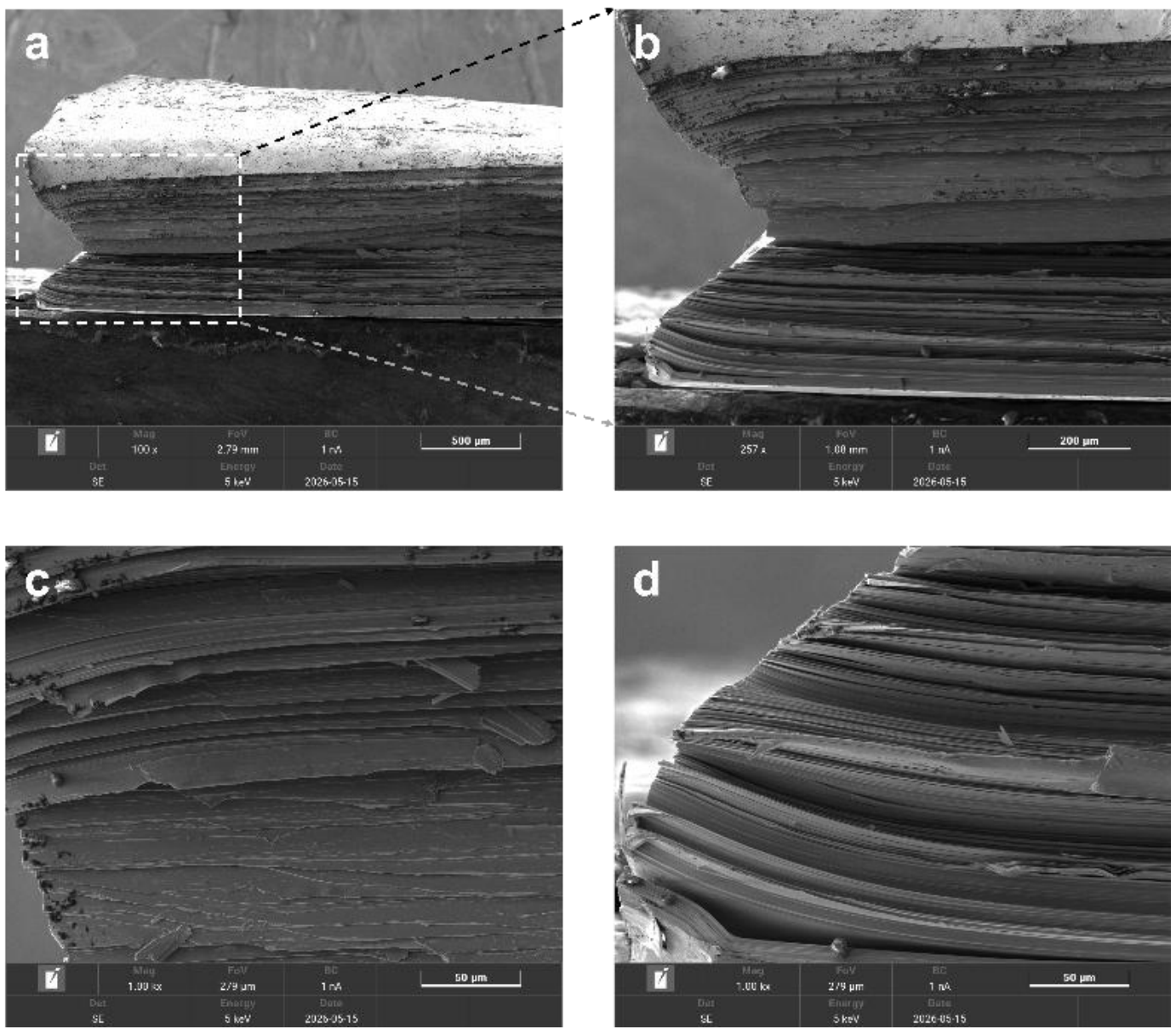

**Figure 3.** SEM image of the fracture surface of GeSe after bending along the *c* axis. The fracture surface is flat and displays typical features of brittle cleavage fracture, with no obvious crack deflection or plastic deformation, indicating brittleness along the *c*-axis (armchair) direction.

An intriguing and thought-provoking observation in the study of GeSe mechanics is that, when GeSe is fabricated into micropillars and subjected to micropillar compression, its mechanical response changes from the plastic deformation observed under macroscopic compression to brittle fracture. The load–depth curve in Fig. 4a rises approximately linearly during initial loading. After the load reaches approximately 8000 μN, the curve exhibits multiple pronounced load drops and serrated fluctuations, rather than the continuously rising strain-hardening plateau observed in the macroscopic compression curve. This indicates abrupt crack initiation and propagation or lamellar spalling during deformation. The in situ SEM sequence (I–IX) in Fig. 4b directly captures this process. The micropillar initially undergoes limited compression under the indenter, after which lamellar delamination, transverse cracks, and fragment sliding and detachment appear at its top. Ultimately, the pillar fails by fragmentation, without exhibiting the large plastic deformation corresponding to approximately 23% strain that the macroscopic sample can sustain. This contrast between macroscopic plasticity and microscopic brittleness indicates that the mechanical behavior of GeSe depends not only on crystallographic orientation and loading direction but also strongly on sample size, defect density, and the state of stress constraint. Under macroscopic compression, weak interlayer van der Waals interactions and interlayer sliding can accommodate large deformation. At the micropillar scale, however, the small sample volume, fewer defects, and strong constraints at the end faces facilitate rapid crack propagation, while interlayer sliding is difficult to activate fully, resulting in brittle behavior.

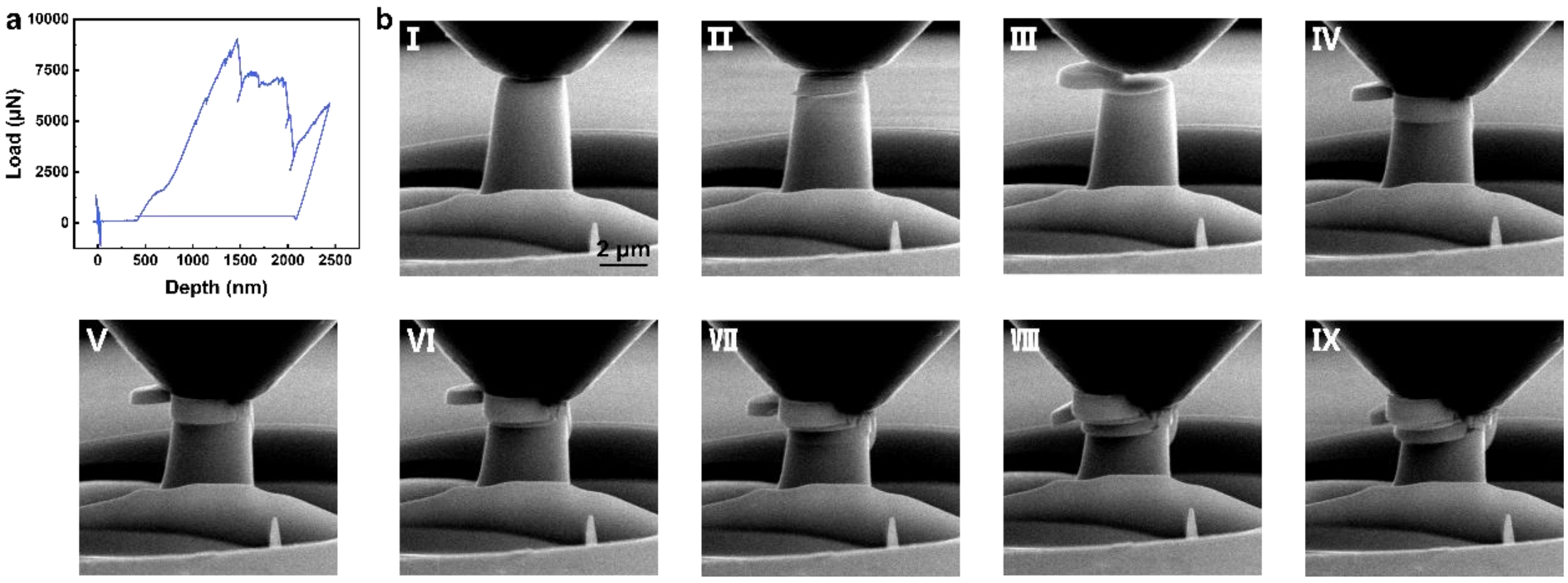

**Figure 4.** Mechanical behavior of GeSe during micropillar compression. (a) Load–depth curve for micropillar compression. (b) In situ SEM sequence (I–IX) showing limited compression, lamellar delamination, transverse cracking, and fragment sliding and detachment during compression, followed by failure through fragmentation. The micropillar has a top diameter of approximately 2.67 μm and a height of approximately 4.57 μm.

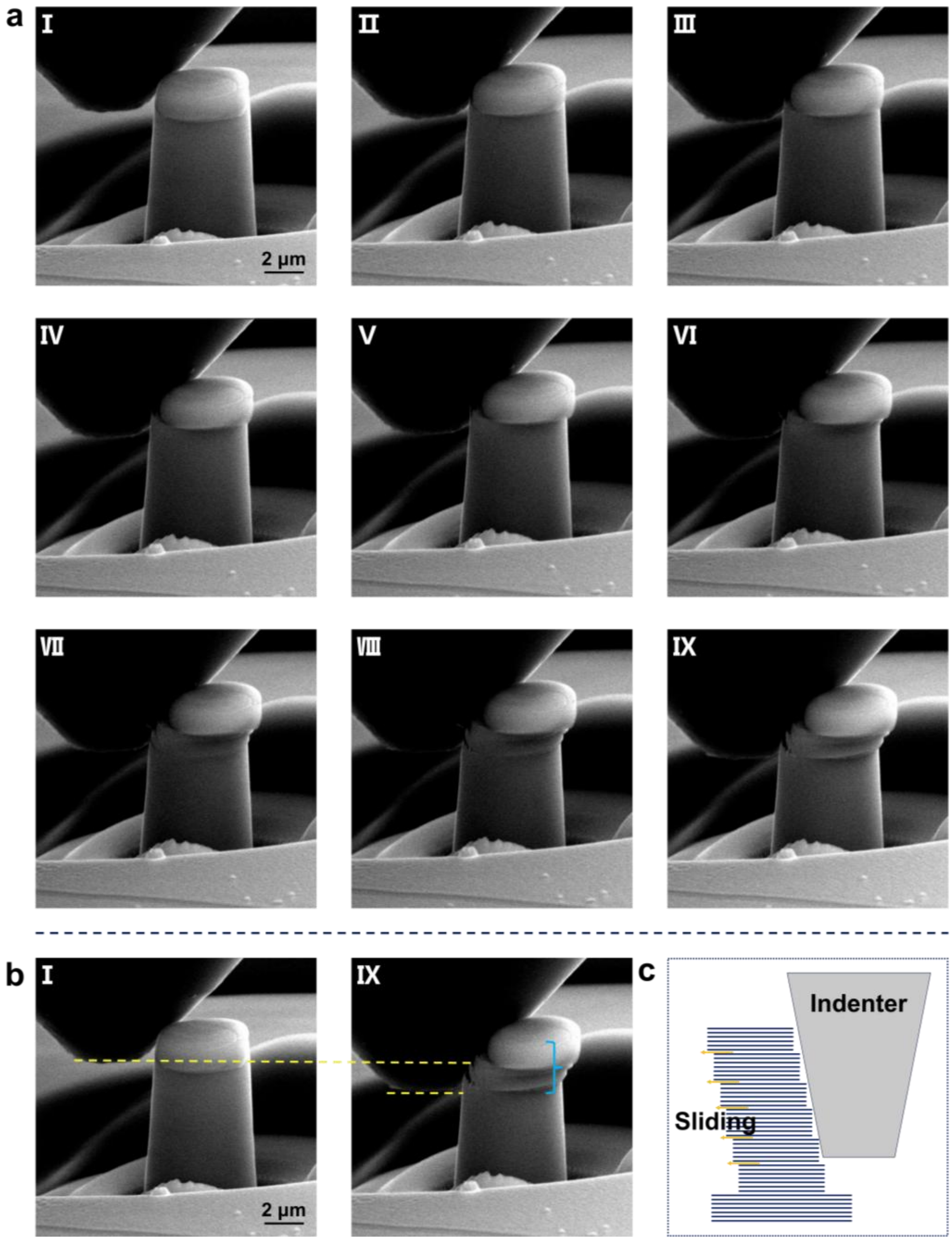

**Figure 5.** Eccentric compression behavior of a GeSe micropillar. (a) In situ SEM sequence (I–IX) of eccentric micropillar compression, showing initial limited compression, followed by offsets between lamellae and lateral sliding in the upper part of the micropillar, and ultimately pronounced interlayer sliding rather than the overall fragmentation observed under conventional micropillar compression. (b) Comparison between states I and IX. Because the indenter acts below the top surface of the micropillar, the indenter displacement indicated by the yellow dashed lines is smaller than the actual interaction distance between the indenter and the micropillar. (c) Schematic illustration of eccentric compression.

We further performed eccentric micropillar compression experiments to examine the effect of a shear stress component on the deformation behavior of GeSe. As shown in Fig. 5a, the indenter did not load the micropillar directly from above but acted on one side below its top surface, producing eccentric compression. During initial loading (I–III), the upper portion of the micropillar underwent limited compression. As the indenter moved further downward (IV–VI), distinct offsets between lamellae and lateral sliding developed in the upper portion.

By stages VII–IX, the top lamellae exhibited pronounced sliding and shear deformation along the interlayer direction, while the pillar did not rapidly undergo overall fragmentation as it did under conventional micropillar compression. Figure 5b compares states I and IX. Because the initial indenter was below the top surface of the micropillar, the indenter displacement indicated by the yellow dashed lines is smaller than the actual interaction distance between the indenter and the micropillar. This indicates that the indenter displacement was accommodated not only by axial compression but also by lateral pushing and interlayer sliding. The schematic in Fig. 5c further illustrates that eccentric loading introduces a substantial shear stress component within the micropillar, promoting interlayer sliding of the lamellae. In contrast to the brittle fracture characterized by rapid crack propagation and fragment detachment under conventional micropillar compression, GeSe exhibits pronounced deformation accommodated by interlayer sliding under eccentric micropillar compression. This directly demonstrates that its weak interlayer van der Waals interactions readily enable interlayer sliding. This comparison shows that, even at the same micrometer scale, the loading mode (axial compression versus eccentric shear) and stress state can substantially alter the plastic or brittle behavior of GeSe.

Using GeSe as a representative model, this study systematically reveals how plasticity and ductile–brittle behavior in plastic inorganic semiconductors may depend on direction, scale, and loading mode. Under macroscopic compression perpendicular to the layers, GeSe sustains approximately 23% strain without fracture and exhibits pronounced plasticity. However, it displays typical brittle cleavage fracture during bending along the *c* axis and changes to brittle fragmentation under conventional micropillar compression. Under eccentric micropillar compression, the shear component instead activates interlayer sliding, accommodating deformation. These apparently contradictory results are not isolated phenomena; they reflect the combined control exerted by bonding topology, interlayer van der Waals interactions, defect density, stress constraints, and strain path over the deformation mechanisms of layered plastic inorganic semiconductors. Similar to isostructural materials such as SnSe, GeSe readily undergoes plastic deformation along the *b*-axis (zigzag) direction because of dense covalent bonding and favorable slip paths, whereas it tends to undergo brittle fracture along the *c*-axis (armchair) direction because the lamellae readily cleave. As sample size decreases, defects become fewer, or the stress state changes, the competition between crack propagation and interlayer sliding is reversed, giving rise to the contrast between macroscopic plasticity and microscopic brittleness.

Extending these insights to plastic inorganic semiconductors as a whole suggests that the "plasticity" of plastic inorganic materials cannot be defined by a single criterion based solely on intrinsic material properties. Instead, it represents an overall response that strongly depends on crystallographic orientation, loading mode, sample size, stress state, defect configuration, and test conditions. Therefore, when determining whether a plastic inorganic semiconductor truly exhibits plasticity, one-sided conclusions based on a single direction, length scale, or testing technique must be abandoned. A systematic evaluation framework integrating multiple length scales, directions, loading modes, and characterization techniques must be established, together with unified and rigorous criteria for plasticity. This will help accurately elucidate the deformation mechanisms and structure–property relationships of plastic inorganic semiconductors and provide a more reliable basis for materials design and selection for flexible electronics, wearable devices, curved thermoelectric devices, and other applications requiring specific deformation capabilities. As more plastic material systems and mechanisms are discovered, our understanding of plasticity as jointly governed by direction, scale, and loading mode will become more complete, advancing plastic inorganic semiconductors from fundamental research toward practical applications.